\documentclass[pre,showpacs,superscriptaddress]{revtex4}
\usepackage{graphics,amsmath,amssymb}
\usepackage{graphics}
\usepackage{bm}

\newcommand{\sinp}{\affiliation{Theoretical Condensed Matter Physics Division and Centre for Applied Mathematics and Computational Science,\\Saha Institute of Nuclear Physics, Block-AF, Sector-I Bidhannagar, Kolkata-700064, India.}}
\newcommand{\oxf}{\affiliation{Rudolf Peierls Centre for Theoretical Physics, Oxford University, 1 Keble Road, Oxford, OX1 3NP, UK.}}
\begin{document}

%%%%%%%%%%%%%%%%%%%%%%%%%%%%%%%%%%%%%%%%%%%%%%%%%%%%%%%%%%%%%
\title{Master equation for a kinetic model of trading market and its 
analytic solution}
%%%%%%%%%%%%%%%%%%%%%%%%%%%%%%%%%%%%%%%%%%%%%%%%%%%%%%%%%%%%%

%%%%%%%%%%%%%%%%%%%%%%%%%%%%%%%%%%%%%%%%%%%%%%%%%%%%%%%%%%%%%
\author{Arnab Chatterjee}
\email{arnab.chatterjee@saha.ac.in}
\sinp
\author{Bikas K. Chakrabarti}
\email{bikask.chakrabarti@saha.ac.in}
\sinp\oxf
\author{Robin B. Stinchcombe}
\email{stinch@thphys.ox.ac.uk}
\oxf\sinp
%%%%%%%%%%%%%%%%%%%%%%%%%%%%%%%%%%%%%%%%%%%%%%%%%%%%%%%%%%%%%

%%%%%%%%%%%%%%%%%%%%%%%%%%%%%%%%%%%%%%%%%%%%%%%%%%%%%%%%%%%%%
\begin{abstract}
We analyze an ideal gas like model of a trading market with quenched random 
saving factors for its agents and show that the steady state income ($m$) 
distribution $P(m)$ in the model has a power law tail with Pareto index $\nu$ 
exactly equal to unity, confirming the earlier numerical studies on this model.
The analysis starts with the development of a master equation for the time
development of $P(m)$. Precise solutions are then obtained in some special
cases.
\end{abstract}
%%%%%%%%%%%%%%%%%%%%%%%%%%%%%%%%%%%%%%%%%%%%%%%%%%%%%%%%%%%%%
\maketitle
%%%%%%%%%%%%%%%%%%%%%%%%%%%%%%%%%%%%%%%%%%%%%%%%%%%%%%%%%%%%%

%%%%%%%%%%%%%%%%%%%%%%%%%%%%%%%%%%%%%%%%%%%%%%%%%%%%%%%%%%%%%%%%%%%%
\section{Introduction}
%%%%%%%%%%%%%%%%%%%%%%%%%%%%%%%%%%%%%%%%%%%%%%%%%%%%%%%%%%%%%%%%%%%%
\noindent
The distribution of wealth among individuals in an economy has been an
important area of research in economics, for more than a hundred
years. Pareto \cite{Pareto:1897} first quantified the high-end of the income
distribution in a society and found it to follow a power-law 
$P(m) \sim m^{-(1+\nu)}$, where $P$ gives the normalized number of people
with income $m$, and the exponent $\nu$, called the Pareto index, was found 
to have a value between 1 and 3.

Considerable investigations with real data during the last ten years revealed
that the tail of the income distribution indeed follows the above
mentioned behavior and the value of the Pareto index $\nu$ is generally 
seen to vary between 1 and 2.5 \cite{Oliveira:1999,realdatag,realdataln}. It is also 
known that typically less than $10 \%$ of the population in any country 
possesses about $40 \%$ of the total wealth of that country and they follow 
the above law. The rest of the low income population, in fact the majority 
($90\%$ or more), follow a different distribution which is debated to be either 
Gibbs \cite{marjit,realdatag} or log-normal \cite{realdataln}.

Much work has been done recently on models of markets, where economic (trading)
activity is analogous to some scattering process 
\cite{marjit,Chakraborti:2000,Chatterjee:2004,Chatterjee:2003,Chakrabarti:2004,othermodels,Slanina:2004}.
We put our attention to models where introducing a saving factor for the 
agents, a wealth distribution similar to that in the real economy can be 
obtained \cite{Chakraborti:2000,Chatterjee:2004}. Savings do play an important role in 
determining the nature of the wealth distribution in an economy and this 
has already been observed in some recent investigations \cite{Willis:2004}.
Two variants of the model have been of recent interest; namely, where the agents
have the same fixed saving factor \cite{Chakraborti:2000}, and where the agents
have a quenched random distribution of saving factors \cite{Chatterjee:2004}.
While the former has been understood to a certain extent (see e.g, 
\cite{Das:2003,Patriarca:2004}), and argued to resemble a gamma distribution 
\cite{Patriarca:2004}, 
attempts to analyze the latter model are still incomplete (see however, 
\cite{Repetowicz:2004}). Further numerical studies \cite{Ding:2003} of time correlations in
the model seem to indicate even more intriguing features of the model.
In this paper, we intend to analyze the second market model with randomly
distributed saving factor, using a master equation type approach similar
to kinetic models of condensed matter.

%%%%%%%%%%%%%%%%%%%%%%%%%%%%%%%%%%%%%%%%%%%%%%%%%%%%%%%%%%%%%%%%%%%%
\section{The Model}
%%%%%%%%%%%%%%%%%%%%%%%%%%%%%%%%%%%%%%%%%%%%%%%%%%%%%%%%%%%%%%%%%%%%
The market consists of $N$ (fixed) agents, each having money $m_i(t)$ at
time $t$ ($i=1,2,\ldots,N$). The total money $M$ ($=\sum_i^N m_i(t)$) in the
market is also fixed. Each agent $i$ has a saving factor $\lambda_i$ 
($0 \le \lambda_i < 1$) such
that in any trading (considered as a scattering) the agent saves a fraction 
$\lambda_i$ of its money $m_i(t)$ at that time and offers the rest 
$(1-\lambda_i)m_i(t)$ for random trading. We assume each trading to be
a two-body (scattering) process. The evolution of money in such a trading
can be written as:
\begin{equation}
\label{mi}
m_i(t+1)=\lambda_i m_i(t) + \epsilon_{ij} \left[(1-\lambda_i)m_i(t) + (1-\lambda_j)m_j(t)\right], 
\end{equation}
\begin{equation}
\label{mj}
m_j(t+1)=\lambda_j m_j(t) + (1-\epsilon_{ij}) \left[(1-\lambda_i)m_i(t) + (1-\lambda_j)m_j(t)\right]
\end{equation}

\noindent
where each $m_i \ge 0$ and $\epsilon_{ij}$ is a random fraction 
($0 \le \epsilon \le 1$). Typical numerical results for the steady state money
distribution in such a model is shown in Fig. 1(a) for uniform distribution of
$\lambda_i$ ($0 \le \lambda_i < 1$) among the agents.

%%%%%%%%%%%%%%%%%%%%%%%%%%%%%%%%%%%%%%%%%%%%%%%%%%%%%%%%%%%%%%%%%
%%%%%%%%FIGURE%%%%%%%%%%%%%%%%%%%%%%%%%%%%%%%%%%%%%%%%%%%%%%%%%%%
\begin{figure}
\centering{
\resizebox*{8.3cm}{!}{\includegraphics{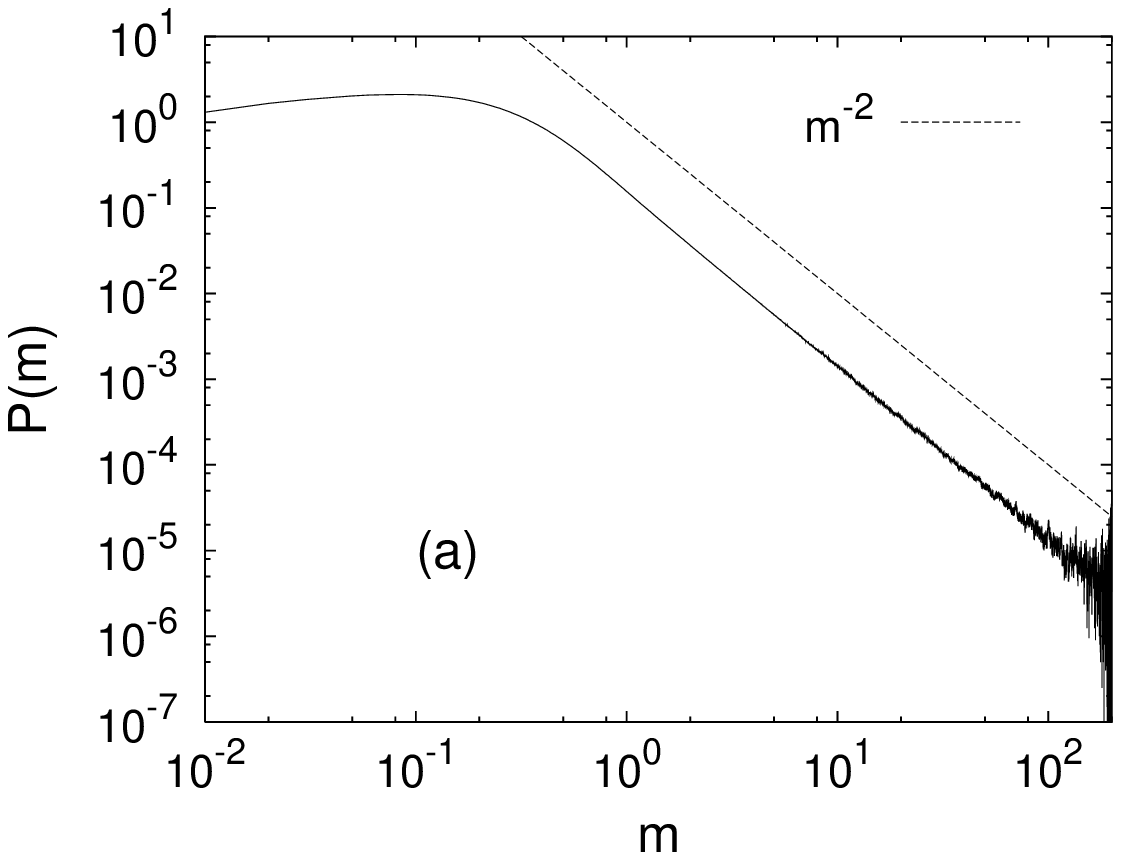}}
\resizebox*{8.3cm}{!}{\includegraphics{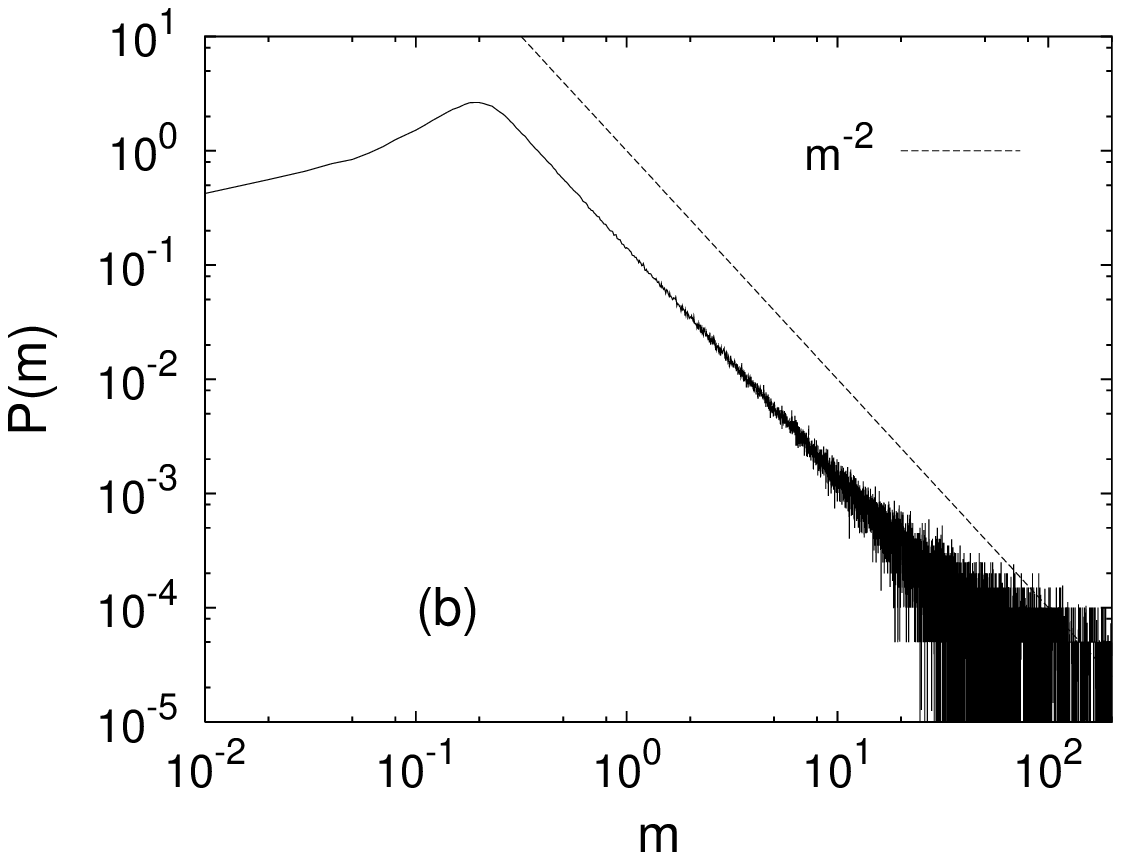}}
}
\caption{\label{latt}
Steady state money distribution $P(m)$ against $m$ in a numerical simulation
of a market with $N=200$, following equations (\ref{mi}) and (\ref{mj}) 
with (a) $\epsilon_{ij}$ randomly distributed in the interval $0$ to $1$
and (b) $\epsilon_{ij}=1/2$.
The dotted lines correspond to $m^{-(1+\nu)}$; $\nu=1$.
}
\end{figure}
%%%%%%%%FIGURE%%%%%%%%%%%%%%%%%%%%%%%%%%%%%%%%%%%%%%%%%%%%%%%%%%%
%%%%%%%%%%%%%%%%%%%%%%%%%%%%%%%%%%%%%%%%%%%%%%%%%%%%%%%%%%%%%%%%%

%%%%%%%%%%%%%%%%%%%%%%%%%%%%%%%%%%%%%%%%%%%%%%%%%%%%%%%%%%%%%%%%%%%%
\section{Dynamics of money exchange}
%%%%%%%%%%%%%%%%%%%%%%%%%%%%%%%%%%%%%%%%%%%%%%%%%%%%%%%%%%%%%%%%%%%%
We will now investigate the steady state distribution of money resulting
from the above two equations representing the trading and money dynamics.
We will now solve the dynamics of money distribution in two limits. In one
case, we study the evolution of the mutual money difference among the agents 
and look for a self-consistent equation for its steady state distribution.
In the other case, we develop a master equation for the money distribution
function.

%%%%%%%%%%%%%%%%%%%%%%%%%%%%%%%%%%%%%%%%%%%%%%%%%%%%%%%%%%%%%%%%%%%%
\subsection{Distribution of money difference}
%%%%%%%%%%%%%%%%%%%%%%%%%%%%%%%%%%%%%%%%%%%%%%%%%%%%%%%%%%%%%%%%%%%%
Clearly in the process as considered above, the total money $(m_i+m_j)$ 
of the pair of agents $i$ and $j$ remains
constant, while the difference $\Delta m_{ij}$ evolves as
\begin{equation}
\label{dDelmijtt}
(\Delta m_{ij})_{t+1} \equiv (m_i-m_j)_{t+1} =
\left( \frac{\lambda_i+\lambda_j}{2} \right)(\Delta m_{ij})_t +
\left( \frac{\lambda_i-\lambda_j}{2} \right)(m_i+m_j)_t +
(2 \epsilon_{ij} -1)[(1-\lambda_i)m_i(t)+(1-\lambda_j)m_j(t)].
\end{equation}

\noindent
Numerically, as shown in Fig. 1, we observe that the steady state money 
distribution in the market becomes a power law, following such tradings 
when the saving factor $\lambda_i$ of the agents remain constant over time 
but varies from agent to agent widely. As shown in the numerical simulation 
results for $P(m)$ in Fig. 1(b), the law, as well as the exponent, remains 
unchanged even when $\epsilon_{ij}=1/2$ for every trading. 
This can be justified by the earlier numerical observation 
\cite{Chakraborti:2000,Chatterjee:2004} for fixed $\lambda$ market 
($\lambda_i = \lambda$ for 
all $i$) that in the steady state, criticality occurs as $\lambda \to 1$
where of course the dynamics becomes extremely slow. In other words,
after the steady state is realized, the third term in (\ref{dDelmijtt})
becomes unimportant for the critical behavior.
We therefore concentrate on this case, where the above evolution equation for 
$\Delta m_{ij}$ can be written in a more simplified form as
\begin{equation}
\label{dDelmt}
(\Delta m_{ij})_{t+1} = \alpha_{ij}(\Delta m_{ij})_t + \beta_{ij}(m_i+m_j)_t,
\end{equation}

\noindent
where $\alpha_{ij}=\frac{1}{2}(\lambda_i+\lambda_j)$ and 
$\beta_{ij}=\frac{1}{2}(\lambda_i-\lambda_j)$. As such, $0 \le \alpha < 1$ and
$-\frac{1}{2} < \beta < \frac{1}{2}$. 

The steady state probability distribution $D$ for the modulus 
$\Delta = |\Delta m|$ of the mutual money difference between any two agents 
in the market can be obtained from (\ref{dDelmt}) in the following way
provided $\Delta$ is very much larger than the average money per agent $=M/N$.
%This is because, large $\Delta$ can appear from `scattering' involving
%$m_i - m_j = \pm \Delta$ and when either $m_i$ or $m_j$ is small. When both
%$m_i$ and $m_j$ are large, maintaining a large $\Delta$ between them,
%their probability is much smaller and hence their contribution.
This is because, using eqn. (\ref{dDelmt}), large $\Delta$ can appear at 
$t+1$, say, from `scattering' from any situation at $t$ for which the right 
hand side of eqn. (\ref{dDelmt}) is large. The possibilities are (at $t$)  
$m_i$ large (rare) and $m_j$ not large, where the right hand side of eqn. 
(\ref{dDelmt}) becomes $\sim (\alpha_{ij} + \beta_{ij})(\Delta_{ij})_t$; 
or  $m_j$ large (rare) and $m_i$ not large (making the right hand side of eqn. 
(\ref{dDelmt}) becomes $\sim (\alpha_{ij} - \beta_{ij})(\Delta_{ij})_t$); 
or when $m_i$ and $m_j$ are both large, which is a much rarer situation than 
the first two and hence is negligible.
Then if, say, $m_i$ is large and $m_j$ is not, the right hand 
side of (\ref{dDelmt}) becomes $\sim (\alpha_{ij}+ \beta_{ij})(\Delta_{ij})_t$
and so on. Consequently for large $\Delta$ the distribution $D$ satisfies
\begin{eqnarray}
\label{DDel}
D(\Delta)
&=& \int d \Delta^\prime \; D(\Delta^\prime) \;
\langle 
\delta (\Delta -(\alpha + \beta) \Delta^\prime) +
\delta (\Delta -(\alpha - \beta) \Delta^\prime) 
\rangle \nonumber\\
&=& 
2 \langle 
\left( \frac{1}{\lambda} \right)
\; D
\left( \frac{\Delta}{\lambda} \right)
\rangle,
\end{eqnarray}

\noindent
where we have used the symmetry of the $\beta$ distribution and the relation 
$\alpha_{ij} + \beta_{ij}=\lambda_i$, and have suppressed labels $i$, $j$. 
Here $\langle \ldots \rangle$ denote average over $\lambda$ distribution
in the market. 
Taking now a uniform random distribution of the saving factor $\lambda$, 
$\rho(\lambda) = 1$ for $0 \le \lambda < 1$, and assuming 
$D(\Delta) \sim \Delta^{-(1+\gamma)}$ for large $\Delta$, we get
\begin{equation}
\label{gammaex}
1=2 \int d \lambda \; \lambda^\gamma = 2 (1+\gamma)^{-1},
\end{equation}

\noindent
giving $\gamma=1$.
No other value fits the above equation. This also
indicates that the money distribution $P(m)$ in the market also follows a 
similar power law variation, $P(m) \sim m^{-(1+\nu)}$ and $\nu=\gamma$.
We will now show in a more rigorous way that indeed the only stable solution 
corresponds to $\nu=1$, as observed numerically 
\cite{Chakrabarti:2004,Chatterjee:2003,Chatterjee:2004}.

%%%%%%%%%%%%%%%%%%%%%%%%%%%%%%%%%%%%%%%%%%%%%%%%%%%%%%%%%%%%%%%%%%%%
\subsection{Master equation and its analysis}
%%%%%%%%%%%%%%%%%%%%%%%%%%%%%%%%%%%%%%%%%%%%%%%%%%%%%%%%%%%%%%%%%%%%
We now proceed to develop a Boltzmann-like master equation for the time
development of $P(m,t)$, the probability distribution of money in the
market. We again consider the case $\epsilon_{ij}=\frac{1}{2}$ in 
(\ref{mi}) and (\ref{mj}) and rewrite them as
\begin{equation}
\label{Amat1}
\left( 
\begin{array}{c}
m_i\\m_j
\end{array}
\right)_{t+1}
= \mathcal{A}
\left( 
\begin{array}{c}
m_i\\m_j
\end{array}
\right)_t
\end{equation}

\noindent
where 
\begin{equation}
\label{Amu}
\mathcal{A}=
\left( 
\begin{array}{cc}
\mu_i^+ & \mu_j^-\\
\mu_i^- & \mu_j^+
\end{array}
\right);\quad
\mu^\pm = \frac{1}{2} (1 \pm \lambda).
\end{equation}

\noindent
Collecting the contributions from terms scattering in and subtracting
those scattering out, we can
write the master equation for $P(m,t)$ as (cf. \cite{Slanina:2004})
\begin{eqnarray}
\label{mstr}
P(m,t+\Delta t) - P(m,t)
&=&
\langle
\int dm_i \int dm_j \; P(m_i,t) P(m_j,t) \nonumber\\
&& \times
\left\{
\left[
\delta (\{\mathcal{A} \;{\bf m}\}_i -m)+
\delta (\{\mathcal{A} \;{\bf m}\}_j -m)
\right]
-
\left[
\delta (m_i-m)+
\delta (m_j-m)
\right]
\right\}
\rangle 
\nonumber\\
& = & 
\langle 
\int dm_i \int dm_j \; P(m_i,t)P(m_j,t) \nonumber\\
& &
\times
[
\delta (\mu_i^+ m_i + \mu_j^- m_j -m) +
\delta (\mu_i^- m_i + \mu_j^+ m_j -m) -
\delta (m_i -m) + \delta (m_j -m)
]
\rangle.
\end{eqnarray}

\noindent 
The above equation can be rewritten as
\begin{equation}
\label{partial}
\frac{\partial P(m,t)}{\partial t} + P(m,t) =
\langle
\int d m_i \int d m_j \; P(m_i,t)P(m_j,t)\;
\delta(\mu_i^+ m_i + \mu_j^- m_j -m)
\rangle,
\end{equation}

\noindent
which in the steady state gives
\begin{equation}
\label{aftrpartial}
P(m) =
\langle
\int d m_i \int d m_j \;P(m_i)P(m_j)\;
\delta(\mu_i^+  m_i +\mu_j^- m_j -m)
\rangle.
\end{equation}

Writing $m_i \mu_i^+=xm$, we can decompose the range $[0,1]$ of $x$ 
into three regions:
$[0,\kappa]$, $[\kappa,1-\kappa^\prime]$ and $[1-\kappa^\prime,1]$.
Collecting the relevant terms in the three regions, we can rewrite the equation
for $P(m)$ above as
\begin{eqnarray}
\label{mumu}
P(m)&=&
\langle
\frac{m}{\mu^+ \mu^-}
\int_0^1 dx
P \left( \frac{xm}{\mu^+} \right)
P \left( \frac{m(1-x)}{\mu^-} \right)
\rangle \nonumber\\
&=&
\langle
\frac{m}{\mu^+ \mu^-}
\left\{ 
P \left( \frac{m}{\mu^-} \right) \frac{\mu^+}{m}
\int_0^{\frac{\kappa m}{\mu^+}} dy P(y) +
P \left( \frac{m}{\mu^+} \right) \frac{\mu^-}{m}
\int_0^{\frac{\kappa^\prime m}{\mu^-}} dy P(y) +
\int_\kappa^{1-\kappa^\prime}
dx P \left( \frac{xm}{\mu^+} \right)
P \left( \frac{m(1-x)}{\mu^-} \right)
\right\}
\rangle \nonumber \\
&&
\end{eqnarray}

\noindent
where the result applies for $\kappa$ and $\kappa^\prime$ sufficiently small.
If we take $m \gg 1/\kappa$, $m \gg 1/\kappa^\prime$
and $\kappa,\kappa^\prime \rightarrow 0$ ($m \rightarrow \infty$),
then
\begin{equation}
\label{mumu1}
P(m)=\langle
\frac{m}{\mu^+ \mu^-}
\left\{
P \left( \frac{m}{\mu^-} \right) \frac{\mu^+}{m} +
P \left( \frac{m}{\mu^+} \right) \frac{\mu^-}{m} +
\int_\kappa^{1-\kappa^\prime}
dx P \left( \frac{xm}{\mu^+} \right)
P \left( \frac{m(1-x)}{\mu^-} \right)
\right\}
\rangle.
\end{equation}

\noindent
Assuming now as before, $P(m) = A/m^{1+\nu}$ for 
$m \rightarrow \infty$, we get
\begin{equation}
\label{mugama}
1
= 
\langle
(\mu^+)^\nu + (\mu^-)^\nu
\rangle
\equiv
\int \int d\mu^+ d\mu^- p(\mu^+) q(\mu^-)
\left[ 
(\mu^+)^\nu + (\mu^-)^\nu
\right],
\end{equation}

\noindent
as the ratio of the third term in (\ref{mumu1}) to the other terms vanishes like
$(m \kappa)^{-\nu}$, $(m \kappa^\prime)^{-\nu}$ in this limit and 
$p(\mu^+)$ and $q(\mu^-)$ are the distributions of the variables $\mu^+$ 
and $\mu^-$, which vary uniformly in the ranges $[\frac{1}{2},1]$ and 
$[0,\frac{1}{2}]$ respectively (cf. eqn (\ref{Amu})).
The $i,j$ indices, for $\mu^+$ and $\mu^-$ are again suppressed here in
(\ref{mugama}) and we utilise the fact that $\mu_i^+$ and $\mu_j^-$ are
independent for $i \ne j$.
An alternative way of deriving Eqn. (\ref{mugama}) from 
Eqn. (\ref{aftrpartial}) is to consider the dominant terms 
($\propto x^{-r}$ for $r>0$, or $\propto \ln (1/x)$ for $r=0$) in the
$x \to 0$ limit of the integral 
$\int_0^\infty m^{(\nu + r)} P(m) \exp (-mx) dm$ (see Appendix A).
We therefore get from Eqn. (\ref{mugama}), after integrations, 
$1=2/(\nu + 1)$, giving $\nu=1$.

%%%%%%%%%%%%%%%%%%%%%%%%%%%%%%%%%%%%%%%%%%%%%%%%%%%%%%%%%%%%%%%%%%%%
\section{Summary and discussions}
%%%%%%%%%%%%%%%%%%%%%%%%%%%%%%%%%%%%%%%%%%%%%%%%%%%%%%%%%%%%%%%%%%%%
In our models 
\cite{Chakraborti:2000,Chatterjee:2004,Chatterjee:2003,Chakrabarti:2004}, 
we consider the 
ideal-gas-like trading markets where each agent is identified with a gas 
molecule and each trading as an elastic or money-conserving (two-body) 
collision. Unlike in a gas, we introduce a saving factor $\lambda$ for each
agents. Our model, without savings ($\lambda=0$), obviously yield a Gibbs
law for the steady-state money distribution. 
Our numerical results for various widely distributed (quenched) saving
factor $\lambda$ showed \cite{Chatterjee:2004,Chatterjee:2003,Chakrabarti:2004}
that the steady state
income distribution $P(m)$ in the market has a power-law tail
$P(m) \sim m^{-(1+\nu)}$ for large income limit, where $\nu \simeq 1.0$.
This observation has been confirmed in several later numerical studies as well
\cite{Repetowicz:2004,Ding:2003}. 
Since $Q(m) = \int_m^\infty P(m) dm$ can be identified with the inverse
rank, our observation in the model with $\nu = 1$ suggests that the rank of
any agent goes inversely with his/her income/wealth, fitting very well with
the Zipf's original observation \cite{Zipf:1949}.
It has been noted from these numerical 
simulation studies that the large income group people usually have larger 
saving factors \cite{Chatterjee:2004}. This, in fact, compares well with 
observations in real markets \cite{Willis:2004,Dynan:2004}. The time 
correlations induced by the random saving factor also has an interesting 
power-law behavior \cite{Ding:2003}. A master equation for
$P(m,t)$, as in (\ref{mstr}), for the original case (eqns. (\ref{mi}) and 
(\ref{mj})) was first formulated for fixed $\lambda$ ($\lambda_i$ same for all 
$i$), in \cite{Das:2003} and solved numerically. Later, a generalized master 
equation for the same, where $\lambda$ is distributed, was formulated 
and solved in \cite{Repetowicz:2004}.

We have formulated here a Boltzmann-type master equation for the distributed
saving factor case in (\ref{mi}) and (\ref{mj}). Based on the observation that
even in the case with $\epsilon=1/2$ (with $\lambda$ distributed in the range
$0 \le \lambda_i < 1$, $\lambda_i \ne \lambda_j$), in 
(\ref{mi}) and (\ref{mj}),
the steady state money distribution has the same power-law behavior as in the
general case and shows the same Pareto index,
we solve the master equation for this special case. We show that the analytic
results clearly support the power-law for $P(m)$ with the exponent value
$\nu=1$.
Although our analysis of the solution of the master equation is for a 
special case and it cannot be readily extended to explore the wide 
universality of the Pareto exponent as observed in the numerical simulations 
of the various versions of our model \cite{Chatterjee:2004,Repetowicz:2004}, 
let alone the quasi-universality for other $\nu$ values as observed in the 
real markets \cite{Oliveira:1999,realdatag,realdataln},
the demonstration here that the master equation admits of a 
Pareto-like power law solution (for large $m$) with $\nu = 1$, 
should be significant.

Apart from the intriguing observation that Gibbs (1901) and Pareto (1897) 
distributions fall in the same category of models and can appear naturally 
in the century-old and well-established kinetic theory of gas, our study 
indicates the appearance of self-organized criticality in the simplest 
(gas-like) models so far, when the stability effect of savings is incorporated. 
This remarkable effect can be analyzed in terms of master equations 
developed here and can also be studied analytically in the special limits 
considered.

%%%%%%%%%%%%%%%%%%%%%%%%%%%%%%%%%%%%%%%%%%%%%%%%%%%%%%%%%%%%%%%%%%%%%%%
\begin{acknowledgments}
BKC is grateful to the INSA-Royal Society Exchange Programme for financial 
support to visit the Rudolf Peierls Centre for Theoretical Physics, 
Oxford University, UK and
RBS acknowledges EPSRC support under the grants GR/R83712/01 and GR/M04426
for this work and wishes to thank the Saha Institute of Nuclear Physics
for hospitality during a related visit to Kolkata, India.
\end{acknowledgments}
%%%%%%%%%%%%%%%%%%%%%%%%%%%%%%%%%%%%%%%%%%%%%%%%%%%%%%%%%%%%%%%%%%%%%%%

%%%%%%%%%%%%%%%%%%%%%%%%%%%%%%%%%%%%%%%%%%%%%%%%%%%%%%%%%%%%%%%%%%%%%%%
\appendix
\section{Alternative solution of the steady state master equation (\ref{aftrpartial})}

Let $S_r (x)= \int_0^\infty dm P(m) m^{\nu+r} \exp(-mx)$; $r \ge 0, x > 0$.
If $P(m) = A/m^{1+\nu}$, then
\begin{eqnarray}
\label{app1}
S_r (x)
&=& 
A \int_0^\infty dm \; m^{r-1} \exp(-mx) \nonumber\\
&\sim&
A \; \frac{x^{-r}}{r} \;\;\;\;\;\;\;\;\; {\rm if} \; \; r > 0 \nonumber\\
&\sim&
A \ln \left( \frac{1}{x} \right)\; \;  \; \; {\rm if} \; \; r = 0.
\end{eqnarray}

From eqn. (\ref{aftrpartial}), we can write
\begin{eqnarray}
\label{app2}
S_r(x)
&=&
\langle
\int_0^\infty dm_i \int_0^\infty dm_j \; P(m_i) P(m_j)
(m_i \mu_i^+ + m_j \mu_j^-)^{\nu+r}
\exp[-(m_i \mu_i^+ + m_j \mu_j^-)x]
\rangle \nonumber\\
&\simeq&
\int_0^\infty dm_i \; A m_i^{r-1}
\langle
\exp(-m_i \mu_i^+ x) \left(\mu_i^+ \right)^{\nu + r}
\rangle
\left[
\int_0^\infty dm_j \; P(m_j)
\langle
\exp(-m_j \mu_j^- x) 
\rangle
\right] \nonumber\\
&& +
\int_0^\infty dm_j \; A m_j^{r-1}
\langle
\exp(-m_j \mu_j^- x) \left(\mu_j^- \right)^{\nu + r}
\rangle
\left[
\int_0^\infty dm_i \; P(m_i)
\langle
\exp(-m_i \mu_i^+ x) 
\rangle
\right]
\end{eqnarray}

\noindent
or,
\begin{eqnarray}
\label{app3}
S_r(x)
&=&
\int_{\frac{1}{2}}^1 d\mu_i^+ \; p(\mu_i^+)
\left( 
\int_0^\infty d m_i \; A m_i^{r-1} \exp(-m_i \mu_i^+ x)
\right)
\left( 
\mu_i^+
\right)^{\nu+r} \nonumber \\
&& +
\int_0^{\frac{1}{2}} d\mu_j^- \; q(\mu_j^-)
\left( 
\int_0^\infty d m_j \; A m_j^{r-1} \exp(-m_j \mu_j^- x)
\right)
\left( 
\mu_j^-
\right)^{\nu+r},
\end{eqnarray}

\noindent
since for small $x$, the terms in the square brackets in (\ref{app2})
approach unity. We can therefore rewrite (\ref{app3}) as
\begin{equation}
\label{app4}
S_r (x) = 2 \left[
\int_{\frac{1}{2}}^1 d \mu^+ (\mu^+)^{\nu + r} S_r (x \mu^+)
+
\int_0^{\frac{1}{2}} d \mu^- (\mu^-)^{\nu + r} S_r (x \mu^-)
\right].
\end{equation}

Using now the forms of $S_r(x)$ as in (\ref{app1}), and collecting terms
of order $x^{-r}$ (for $r>0$) or of order $\ln (1/x)$ (for $r=0$)
from both sides of (\ref{app4}), we get (\ref{mugama}).

%%%%%%%%%%%%%%%%%%%%%%%%%%%%%%%%%%%%%%%%%%%%%%%%%%%%%%%%%%%%%%%%%%%%%%%
%%%%%%%%%%%%%%%%%%%%%%%%%%%%%%%%%%%%%%%%%%%%%%%%%%%%%%%%%%%%%%%%%%%%%%%

%%%%%%%%%%%%%%%%%%%%%%%%%%%%%%%%%%%%%%%%%%%%%%%%%%%%%%%%%%%%%%%%%%%%


\begin{thebibliography}{1}

\bibitem{Pareto:1897}
V. Pareto, \emph{Cours d'economie Politique} (F. Rouge, Lausanne, 1897);
\emph{Econophysics of Wealth Distributions}, Eds. A. Chatterjee, S.
Yarlagadda, and B. K. Chakrabarti (Springer-Verlag Italia, Milan, 2005).

\bibitem{Oliveira:1999}
S. Moss de Oliveira, P. M. C. de Oliveira and D. Stauffer,
\emph{Evolution, Money, War and Computers} (B. G. Tuebner, Stuttgart, Leipzig,
1999).

\bibitem{realdatag}
M. Levy and S. Solomon, Physica A {\bf 242} (1997) 90;
A. A. Dr\u{a}gulescu and V. M. Yakovenko, Physica A {\bf 299} (2001) 213; 
H. Aoyama,  W. Souma and Y. Fujiwara, Physica A {\bf 324} (2003) 352.

\bibitem{realdataln}
T. Di Matteo, T. Aste, and S. T. Hyde in \emph{The Physics of Complex Systems
(New Advances and Perspectives)}, Eds. F. Mallamace and H. E. Stanley
(IOS Press, Amsterdam, 2004) p.435;
F. Clementi and M. Gallegati, Physica A 350 (2005) 427.

\bibitem{marjit}
B. K. Chakrabarti and S. Marjit, Ind. J. Phys. B {\bf 69} (1995) 681;
S. Ispolatov, P. L. Krapivsky and S. Redner, Eur. Phys. J. B {\bf 2} (1998) 267;
A. A. Dr\u{a}gulescu and V. M. Yakovenko, Eur. Phys. J. B {\bf 17} (2000) 723.

\bibitem{Chakraborti:2000}
A. Chakraborti and B. K. Chakrabarti, Eur. Phys. J. B {\bf 17} (2000) 167.

\bibitem{Chatterjee:2004}
A. Chatterjee, B. K. Chakrabarti and S. S. Manna, Physica A {\bf 335}
(2004) 155.

\bibitem{Chatterjee:2003}
A. Chatterjee, B. K. Chakrabarti and S. S. Manna, Phys. Scr. T {\bf 106} (2003) 36.

\bibitem{Chakrabarti:2004}
B. K. Chakrabarti and A. Chatterjee, in \emph{Application of Econophysics},
Proc. 2nd Nikkei Econophys. Symp. (Tokyo, 2002), ed. H. Takayasu, 
(Springer, Tokyo, 2004), pp. 280-285, cond-mat/0302147.

\bibitem{othermodels}
B. Hayes, Am. Scientist {\bf 90} (Sept-Oct, 2002) 400;
S. Sinha, Phys. Scr. T {\bf 106} (2003) 59;
J. C. Ferrero, Physica A {\bf 341} (2004) 575;
J. R. Iglesias, S. Gon\c{c}alves, G. Abramson and J. L. Vega, Physica A {\bf 342}
(2004) 186;
N. Scafetta, S. Picozzi and B. J. West, Physica D {\bf 193} (2004) 338.

\bibitem{Slanina:2004}
F. Slanina, Phys. Rev. E {\bf 69} (2004) 046102.

\bibitem{Willis:2004}
G. Willis and J. Mimkes, cond-mat/0406694.

\bibitem{Das:2003}
A. Das and S. Yarlagadda, Phys. Scr. T {\bf 106} (2003) 39.

\bibitem{Patriarca:2004}
M. Patriarca, A. Chakraborti and K. Kaski, Phys. Rev. E {\bf 70} (2004) 016104.

\bibitem{Repetowicz:2004}
P. Repetowicz, S. Hutzler and P. Richmond, cond-mat/0407770.

\bibitem{Ding:2003}
N. Ding, N. Xi and Y. Wang,  Eur. Phys. J. B {\bf 36} (2003) 149.

\bibitem{Zipf:1949}
G. K. Zipf, \emph {Human Behavior and the Principle of Least Effort} 
(Addison-Wesley, 1949).

\bibitem{Dynan:2004}
K. E. Dynan, J. Skinner and S. P. Zeldes, J. Pol. Econ. {\bf 112} (2004) 397.

\end{thebibliography}
\end{document}